\documentclass[12pt]{article}
\usepackage{amssymb}

\newcommand{\beq}{\begin{equation} }
\newcommand{\eeq} {\end{equation} }

\begin{document}

\begin{titlepage}
\begin{flushright}
OSU-HEP-02-17\\
hep-ph/0212245\\
December 2002\\
\end{flushright}
\vskip 2cm
\begin{center}
{\large\bf \textbf{Natural $R$--Parity, $\mu$--term, and Fermion Mass Hierarchy\\[0.15in]
From Discrete Gauge Symmetries}}
\vskip 1cm
{\normalsize\bf
K.S.\ Babu\footnote{E-mail address: babu@okstate.edu}, Ilia Gogoladze\footnote{
On leave of absence from: Andronikashvili Institute of Physics, GAS, 380077 Tbilisi,
Georgia.  \\ E-mail address: ilia@hep.phy.okstate.edu}  and
Kai Wang\footnote{E-mail address: wk@okstate.edu}} \\
\vskip 0.5cm
{\it Department of Physics, Oklahoma State University\\
Stillwater, OK~~74078, USA\\ [0.1truecm]
}

\end{center}

\begin{abstract}
In the minimal supersymmetric Standard Model with
seesaw neutrino masses we show how $R$--parity can emerge naturally
as a discrete gauge symmetry.  The same discrete symmetry explains
the smallness of the $\mu$--term (the Higgsino
mass parameter) via the Giudice--Masiero mechanism.  The discrete
gauge anomalies are cancelled by a discrete version of the
Green--Schwarz mechanism.  The simplest symmetry group is found to be
$Z_4$ with a  charge assignment that is compatible with grand
unification.  Several other $Z_N$ gauge symmetries are found for
$N=10,~12,~18,~36$ etc, with some models employing discrete anomaly
cancellation at higher Kac--Moody levels.  Allowing for a flavor
structure in $Z_N$, we show that the same gauge symmetry can also explain
the observed hierarchy in the fermion masses and mixings.
\end{abstract}

\end{titlepage}

\newpage

\section{Introduction}

One of the challenging questions facing supersymmetric extensions of the
Standard Model is an understanding of $R$--parity which is required for
the stability of the proton. In the minimal supersymmetric Standard Model
(MSSM), a discrete $Z_2$ symmetry is usually assumed.  Under this symmetry
the Standard Model (SM) particles are taken to be even while their superpartners
are odd.  The gauge symmetry of MSSM  would allow
baryon number and lepton number violating Yukawa couplings
at the renormalizable level which would result
in rapid proton decay.  The $Z_2$~ $R$--parity forbids such dangerous couplings.

The assumption of $R$--parity has profound implications for supersymmetric particle
search at colliders as well as for cosmology.  At colliders SUSY particles can only
be produced in pairs, and the lightest SUSY particle (LSP), usually a
neutralino, will be stable.  This stable LSP is a leading candidate for
cosmological cold dark matter.

Since $R$--parity is not part of the MSSM gauge symmetry, questions can be raised
about its potential violation arising from quantum gravitational effects.
These effects (associated with worm holes, black holes, etc) are believed
to violate all global symmetries \cite{hawking}.  True gauge symmetries
are however protected from such violations.  When a gauge symmetry breaks
spontaneously, often a discrete subgroup is left intact.  Such discrete
symmetries, called discrete gauge symmetries \cite{kw}, are also immune to
quantum gravitational effects.  Not all discrete symmetries can however be gauge
symmetries.  For instance, since the original continuous gauge symmetry
was free from anomalies, its unbroken discrete subgroup should
be free from discrete gauge anomalies \cite{trivedi,db}.  This imposes a
non--trivial constraint on the surviving discrete symmetry and/or on the
low energy particle content \cite{kw,trivedi,db,ibanez1,ibanez2,maru,ma}.

It will be of great interest to see if $R$--parity of MSSM can be realized
as a discrete gauge symmetry, so that one can rest assured that it wont
be subject to unknown quantum gravitational violations.  This is the
main question we wish to address in this paper.

A seemingly unrelated but equally profound problem facing the MSSM is an
understanding of the origin of the
$\mu$--term, the Higgsino mass parameter.  The $\mu$ parameter is defined through
the superpotential term $W \supset \mu H_u H_d$, where $H_u$ and $H_d$
are the two Higgs doublet superfields of MSSM.  Since the $\mu$--term is
SUSY--preserving and is a singlet of the SM gauge symmetry, its
natural value would seem to be of order the Planck scale. But $\mu
\sim 10^2$ GeV is required for consistent phenomenology.  It will be
desirable, and is often assumed, that the $\mu$ term is related to the
supersymmetry breaking scale.  An attractive scenario which achieves
this is the Guidice--Masiero mechanism \cite{gm} wherein a bare
$\mu$ term in the superpotential  is forbidden by some symmetry.  It
is induced through a higher dimensional Lagrangian term
\begin{equation}
{\cal L} = \int d^4 \theta {H_u H_d Z^* \over M_{\rm Pl}}
\end{equation}
where $Z$ is a spurion field which parametrizes supersymmetry breaking
via $\left \langle F_Z \right \rangle \neq 0$, with
$\left \langle F_Z \right \rangle/M_{\rm Pl} \sim M_{\rm SUSY} \sim 10^2$ GeV.
For this mechanism to work, there must exist a symmetry that forbids a
bare $\mu$ term in the superpotential.  Such a symmetry cannot be a continuous
symmetry, consistent with the requirement of non--zero guaginos masses, and
therefore must be discrete.  It would be desirable to realize this as a discrete
gauge symmetry.

The purpose of this paper is show that it is possible to realize $Z_N$
symmetries as discrete gauge symmetries which act as $R$--parity
and which solve simultaneously  the $\mu$--problem via the Guidice--Masiero
mechanism.  We make use of a discrete version of
the Green--Schwarz mechanism \cite{gs} for anomaly cancellation.
Simple realizations of $R$--parity as a discrete gauge symmetry are
possible which also solve the $\mu$--problem, without enlarging the
particle content of MSSM.  The simplest symmetry group we have found
is a $Z_4$.  Under this $Z_4$ all MSSM matter superfields and the gauginos
carry equal charge of $1$ while the Higgs superfields have zero
charge.  Such a simple charge assignment is
compatible with grand unification.  This charge assignment is anomaly--free
by virtue of the discrete Green--Schwarz mechanism.  Other $Z_N$
symmetries with $N=10,~12,~18,~36,$ etc are also identified, some
realized at higher Kac--Moody level.  Either lepton parity or baryon
parity can be obtained as a discrete symmetry in this approach, with
baryon parity requiring anomaly cancellation at higher Kac--Moody levels.
By allowing for a family--dependent structure in $Z_N$, we show how
it is possible in our framework to explain the observed fermion mass
and mixing hierarchy in a simple way.

Attempts have been made in the past to derive $R$--parity
as a discrete gauge symmetry within MSSM.  Early analyses \cite{ibanez1,
ibanez2} did not incorporate the seesaw mechanism for neutrino mass
or the Guidice--Masiero mechanism for generating the $\mu$--term.
A recent analysis  which includes these features \cite{maru}
has found $Z_9$ and $Z_{18}$ discrete gauge symmetries
as possible candidates for $R$--parity by demanding
these symmetries to be anomaly--free.  It turns out that in these models \cite{maru}
the Kahler potential violates $R$--parity at higher order, leading to cosmologically
disfavored lifetime for the neutralino LSP.  Furthermore, the discrete
charge assignment in these models was not compatible with grand unification with
the MSSM spectrum.
The main difference in our approach is that we make use of the Green--Schwarz
mechanism for discrete anomaly cancellation, which is less restrictive
compared to the straightforward methods.  The outcome differs in several ways, notably in
the realization of
simpler symmetries (eg. $Z_4$), exact $R$--parity
without cosmological problems, and compatibility with grand unification.
It should  be mentioned that enlarging the  particle content of MSSM
has been proposed as a solution to the $R$--parity and $\mu$ problems \cite{ma}.
In contrast to such approaches, in our
framework, the low energy spectrum is identical to
that of MSSM.

This paper is organized as follows.  In Sec. 2 we review briefly the
discrete version of Green--Schwarz anomaly cancellation mechanism.
Sec. 3 contains our main results.
In 3.1 we write down the constraints arising from the Lagrangian of MSSM
and the discrete anomaly cancellation conditions.  In Sec. 3.2
we identify possible discrete gauge symmetries at Kac--Moody level 1
which prevent $R$--parity violating terms.  In Sec. 3.3 we embed these
symmetries to a higher $Z_N$ to solve the $\mu$--problem.
Sec. 3.4 is devoted to solutions based on higher Kac--Moody levels.
In Sec. 4 we show how a simple discrete gauge symmetry can
explain the fermion mass and mixing angle hierarchy.  Finally we conclude
in Sec. 5.

\section{Discrete anomaly cancellation via Green--Schwarz mechanism}

Let us first recall the essence of the Green--Schwartz (GS) anomaly cancellation
mechanism for a $U(1)$ guage symmetry.
String theory when compactified to four dimensions generically contains
an ``anomalous $U(1)_A$" gauge symmetry.  A subset of the gauge anomalies in
the axial vector $U(1)_A$ current can be cancelled via the GS mechanism
in the following way \cite{gs}.
In four dimensions, the Lagrangian for the gauge boson kinetic energy contains
the terms
\begin{equation}
{\cal L}_{\rm kinetic} = \varphi(x) \sum_ik_1 F_i^2 + i\eta (x) \sum_ik_i
F_i \tilde{F_i}
\end{equation}
where $\varphi (x)$ denotes the string dilaton field and $\eta (x)$ is its
axionic partner.  The sum $i$ runs over the different gauge groups in the
model, including $U(1)_A$.  $k_i$ are the Kac--Moody levels for the different
gauge groups, which must be positive integers for the non--Abelian groups,
but may be non--integers for Abelian groups.  The GS mechanism makes use
of the transformation of the string axion field $\eta (x)$ under a $U(1)_A$ gauge
variation $V^\mu_A \rightarrow V^\mu_A + \partial_\mu \theta (x)$,
\begin{equation}
\eta (x) \rightarrow \eta (x) -\theta (x) \delta_{GS}
\end{equation}
where $\delta_{GS}$ is a constant.  If the anomaly coefficients involving
the $U(1)_A$ gauge boson and any other pair of gauge bosons are in the ratio
\begin{equation}
{A_1 \over k_1} = {A_2 \over k_2} = {A_3 \over k_3}, .... = \delta_{GS}~,
\end{equation}
these anomalies will be cancelled by gauge variations of the $U(1)_A$ field
arising from
the second term of Eq. (2).  $\delta_{GS}$ in Eq. (4) is also equal to the
mixed gravitational anomaly, $\delta_{GS} = A_{\rm gravity}/12$ \cite{ramond}.
All other
crossed anomaly coefficients should vanish, since they cannot
be removed by the shift in the string axion field.

Consider the case when the gauge symmetry in four dimensions
just below the string scale is $SU(3)_C \times
SU(2)_L \times U(1)_Y \times U(1)_A$.  Let $A_3$ and $A_2$ denote the anomalies
associated with $[SU(3)_C]^2 \times U(1)_A$ and $[SU(2)_L]^2 \times U(1)_A$
respectively.  Then if $A_3/k_3 = A_2/k_2 = \delta_{GS}$
is satisfied, from Eq. (4), it
follows that these mixed anomalies will be cancelled.  The anomaly in
$[U(1)_Y^2] \times U(1)_A$ can also be cancelled in a similar way if
$A_1/k_1 = \delta_{GS}$.  However, in practice, this last condition is
less useful, since $k_1$ is not constrained to be an integer as
the overall normalization of the hypercharge is
arbitrary.  If the full high energy theory is specified, there can be
constraints on $A_1$ as well.
For example, if hypercharge is embedded into
a simple group such as $SU(5)$ or $SO(10)$, $k_1 = 5/3$ is fixed
since hypercharge is now quantized.  $A_1/k_1 = \delta_{GS}$ will provide
a useful constraint in this case.  We shall remark on this
possibility in our discussions.  Note also that cross anomalies such
as $[SU(3)] \times [U(1)_A]^2$ are automatically zero in the Standard
Model, since the trace of $SU(N)$ generators is zero.  Anomalies of
the type $[U(1)_Y] \times [U(1)_A]^2$ also suffer from the same aribitrariness
from the Abelian levels $k_1$ and $k_A$.  Finally, $[U(1)_A]^3$ anomaly
can be cancelled by the GS mechanism, or by contributions from fields
that are complete singlets of the Standard Model gauge group.

The anomalous $U(1)_A$ symmetry is expected to be broken just below the
string scale.  This occurs when the Fayet--Iliopoulos term associated with
the $U(1)_A$ symmetry is cancelled, so that supersymmetry remains
unbroken near the string scale, by shifting the matter superfields that
carry $U(1)_A$ charges \cite{dsw}.  Although the $U(1)_A$ symmetry is
broken, a $Z_N$ subgroup of $U(1)_A$ can remain intact.  Suppose that we
choose a normalization wherein the $U(1)_A$ charges of all fields are
integers.  (This can be done so long as all the charges are relatively
rational numbers.)  Suppose that the scalar field which acquires a vacuum
expectation value (VEV) and breaks $U(1)_A$ symmetry has a charge $N$
under $U(1)_A$ in this normalization.  A $Z_N$ subgroup is then left unbroken
down to low energies.  We shall identify $R$--parity of MSSM with this
unbroken $Z_N$ symmetry.

The field that acquires a VEV and breaks $U(1)_A$ to $Z_N$ can
supply large masses of order the
string scale to a set of fermions which have Yukawa couplings involving this
field.  Such fields may include Majorana fermions and Dirac fermions.
These heavy fields can carry SM gauge quantum numbers, but they must transform
vectorially under the SM.
In order that their mass terms be invariant under the unbroken $Z_N$, it must be
that
\begin{eqnarray}
2 q_i &=& 0 ~mod~ N ~({\rm Majorana ~fermion}) \nonumber \\
q_i + \bar{q_i} &=& 0 ~mod~N ~({\rm Dirac ~fermion})
\end{eqnarray}
where $q_i$ are the $U(1)_A$ charges of these heavy fermions.  The index
$i$ is a flavor index corresponding to different heavy fields.
These heavy fermions, being chiral under the $U(1)_A$, contribute to
gague anomalies.  Their contribution to the $[SU(3)_C]^2
\times U(1)_A$ gauge anomaly is given by $A_3 =
\sum_i q_i \ell_i = (N/2)\sum_ip_i \ell_i$ (Majorana fermion) or
$A_3 = \sum_i (q_i + \bar{q_i})\ell_i = (N)\sum_ip_i\ell_i$ (Dirac
fermion) where $\ell_i$ is the quadratic index of the relevant fermion under
$SU(3)_C$ and the $p_i$ are integers.
We shall adopt the usual normalization of $\ell = 1/2$ for
fundamental of $SU(N)$.  Then, for the case of heavy Dirac fermion, one has
$A_3 = p (N/2)$ where $p$ is an integer, as the index of
the lowest dimensional (fundamental) representations is 1/2 and those
of all other representations are integer multiples of 1/2.
The same conclusion follows for
the case of Majorana fermions for a slightly different reason.  All real
representations of $SU(3)_C$ (such as an octet) have integer values
of $\ell$, so that $\sum_ip_i \ell_i$ is an integer.  Analogous
conclusions follow for the $[SU(2)_L]^2 \times U(1)_A$ anomaly coefficient.

If the $Z_N$ symmetry that survives to low energies was part of $U(1)_A$,
the $Z_N$ charges of the fermions in the low energy theory must satisfy
a non--trivial condition:  The anomaly coefficients $A_i$ for the full
theory is given by $A_i$ from the low energy sector plus
an integer multiple of $N/2$.  These anomalies should obey Eq. (4),
leading to the discrete version of the Green--Schwarz anomaly cancellation
mechanism:
\begin{equation}
{A_3 + {p_1 N  \over 2} \over k_3} = {A_2 + {p_2 N \over 2} \over k_2} = \delta_{GS}
\end{equation}
with $p_1, ~p_2$ being integers.  Since $\delta_{GS}$ is an unknown constant
(from the effective low energy point of view), the discrete anomaly
cancellation conditions of Eq. (6) are less stringent than those arising
from conventional anomaly cancellations.  If $\delta_{GS} = 0$ in Eq. (6),
the anomaly is cancelled without assistance from the Green--Schwarz mechanism.
We shall not explicitly use the condition that $\delta_{GS} \neq 0$, so
our solutions will contain those obtained by demanding  $\delta_{GS} = 0$ in Eq. (6),
viz., $A_3 = -p_1 (N/2)$, $A_2 = -p_2 (N/2)$ with $p_1,~p_2$ being integers.

In our analysis we shall not explicitly
make use of the condition $A_1/k_1 = A_2/k_2$, since, as mentioned earlier,
the overall normalization of hyperhcarge is arbitrary.  However, once a solution
to the various $Z_N$ charges is obtained, we
can check for the allowed values  $k_1$,
and in particular, if $k_1 = 5/3$ is part of the allowed solutions.
This will be an interesting case for two reasons.  If hypercharge
is embedded in a simple grand unification group such as $SU(5)$, one would
expect $k_1 = 5/3$.  Even without a GUT embedding
$k_1 = 5/3$ is interesting.
We recall that unification of gauge couplings is a necessary
phenomenon in string theory.  Specifically, at tree level, the gauge couplings
of the different gauge groups are related to the string coupling constant
$g_{\rm st}$ which is determined by the VEV of the dilaton field as \cite{ginsparg}
\begin{equation}
k_3 g_3^2 = k_2 g_2^2 = k_1 g_1^2 = g_{\rm st}^2~
\end{equation}
where $k_i$ are the levels of the corresponding Kac--Moody algebra.  In
particular, if $k_1: k_2:k_3 = 5/3:1:1$, we would have
$\sin^2\theta_W = 3/8$ at the string scale, a scenario
identical to that of conventional  gauge coupling unification with simple group
such as $SU(5)$.  For these reasons, we shall pay special attention
to the case $k_1 = 5/3$.

An interesting example of a discrete gauge symmetry
in the MSSM (or the SM) with seesaw neutrino masses is the $Z_6$ subgroup
of $B-L$.  The introduction of the right--handed neutrino for
generating small neutrino masses makes $B-L$ a true gauge symmetry.
When the $\nu^c$ fields acquire super-large Majorana  masses, $U(1)_{B-L}$
breaks down to a discrete $Z_6$ subgroup.  The $Z_6$ charges of the MSSM
fields arising from $B-L$ are displayed in Table 1.  Here we have used
the standard notation for the fermion fields ($Q$ and $L$ being the
left--handed quark and lepton doublets, $u^c,~d^c$ being the quark
singlets, and $e^c,~\nu^c$ being the (conjugates of) the right--handed
electron and the right--handed neutrino singlets).  To obtain the unbroken
$Z_6$ charge, we first multiply te $B-L$ charge by 3 so that they become
integers, then observe that the $\nu^c \nu^c$ Majorana mass term carries
6 units of this integer $B-L$ charge.  Thus this $Z_6$ subgroup
is left unbroken.

It is worth mentioning that the $Z_6$ symmetry
has a $Z_2$ and a $Z_3$ subgroups as well.  In the analysis that follows
in the next section we will be making connections with the $Z_6$ subgroup
of $B-L$ and its $Z_2$ and $Z_3$  subgroups.

\begin{table}[h]
 \begin{center}
  {\renewcommand{\arraystretch}{1.1}
  \begin{tabular}{| c | c  c  c  c  c  c  c  c    |}
     \hline

     \rule[5mm]{0mm}{0pt}
   Field & $Q$ & $u^{c}$ &
   $d^{c}$ & $L$ &
   $e^{c}$ & $\nu^{c}$ &
   $H_{u}$ & $H_{d}$  \\
   \hline
   U(1)$_{B-L}$ & 1/3 & $-1/3$ & $-1/3$ &
   $-1$ & 1 & $1$ & 0 & 0 \\
   \rule[5mm]{0mm}{0pt}
   $Z_6$ & $1$ & 5 & 5 &
   3 & 3 & 3 & 0 & 0 \\
   \hline
  \end{tabular}
  }
  \caption{\footnotesize The $B-L$ charges of the Standard Model fields
  along with the unbroken $Z_6$ subgroup after the neutrino seesaw.}
  \label{b-l}
 \end{center}
\end{table}

Anomalous $U(1)$ symmetry has found applications in addressing the
fermion mass and mixing hierarchy problem \cite{mass}, doublet--triplet
splitting problem in GUT \cite{dvali1}, the strong CP problem \cite{cp},
the $\mu$ problem of SUSY \cite{nilles} and for SUSY breaking \cite{dvali2}.

\section{Discrete  gauge symmetries in the MSSM}

In this section we turn to the identification of discrete gauge symmetries
in the MSSM that can serve as $R$--parity and simultaneously explain the
origin of the $\mu$ term.  We stay with the minimal particle content of MSSM,
with the inclusion of the right--handed neutrinos needed for generating neutrino
masses via the seesaw mechanism \cite{seesaw}.  Anomalies associated with
the discrete gauge symmetry will be cancelled by the Green--Schwarz
mechanism as discussed in Sec. 2.

\subsection{ Constraints from the Lagrangian and discrete anomalies}

We have displayed in Table 2 the particle content of MSSM along with
their charges under an anomalous $U(1)$ gauge symmetry.  The Grassmanian
variable $\theta$ also carries a charge (equal to $\alpha)$, which allows
for the $U(1)$ to be identified as an $R$ symmetry.  $Z$ is a spurion superfield that
acquires a non--zero $F$ component and breaks supersymmetry with
$\left\langle F_Z \right \rangle/M_{\rm Pl} \sim M_{\rm SUSY} \sim 10^2$
GeV.  We shall assume family--independent $U(1)$ symmetry in this section.
Any unbroken discrete symmetry must be family--independent
to be consistent with MSSM phenomenology, that is the reason for focusing on
such symmetries.  In Sec. 4 we shall extend this analysis to flavor--dependent
symmetries, even in that case, we will demand that a flavor--independent
$Z_M$ symmetry is left intact.

\begin{table}[h]
 \begin{center}
  {\renewcommand{\arraystretch}{1.1}
  \begin{tabular}{| c | c  c  c  c  c  c  c  c  c  c |}
     \hline

     \rule[5mm]{0mm}{0pt}
   Field & $Q$ & $u^{c}$ &
   $d^{c}$ & $L$ &
   $e^{c}$ & $\nu^{c}$ &
   $H_{u}$ & $H_{d}$ &
   $\theta$ & $Z$\\
   \hline
   \rule[5mm]{0mm}{0pt}
   SU(3)$_{C}$ &  \bf{3} & $\bf{\bar{3}}$ & $ \bf{\bar{3}}$ &
    \bf{1} &  \bf{1} &  \bf{1} &  \bf{1} &  \bf{1} &  \bf{1} &  \bf{1}\\
    \rule[5mm]{0mm}{0pt}
   SU(2)$_{L}$ &  \bf{2} &  \bf{1} &  \bf{1} &
    \bf{2} &  \bf{1} &  \bf{1} &  \bf{2} &  \bf{2} &  \bf{1} &  \bf{1} \\
    \rule[5mm]{0mm}{0pt}
   U(1)$_{Y}$ & 1/6 & $-2/3$ & 1/3 &
   $-1/2$ & 1 & 0 & 1/2 & $-1/2$ & 0& 0\\
   \rule[5mm]{0mm}{0pt}
   $U(1)_A$ & $q$ & $u$ & $d$ &
   $l$ & $e$ & $n$ & $h$ & $\bar{h}$ & $\alpha$ & $z$\\
   \hline
  \end{tabular}
}
  \caption{\footnotesize The matter superfields of  MSSM along
  with their anomalous $U(1)$ charges. $\theta$ is the Grassmann
  variable.
   $Z$ is the spurion field which is responsible for supersymmetry breaking.}
  \label{MSSM}
 \end{center}
\end{table}

The superpotential of the model, including small neutrino masses via the
seesaw mechanism is
\beq \label{superpotential}
W=Qu^{c}H_u+Qd^{c}H_d+Le^{c}H_d+L\nu^{c}H_u+M_R\nu^{c}\nu^{c},
\eeq
where $M_R$ is the heavy right--handed neutrino
Majorana mass.  We have suppressed Yukawa couplings and generation indices,
which must be understood.

In order to avoid a supersymmetric $\mu$ term in the Lagrangian,
${\cal L}  \supset \mu \int d^2\theta H_u H_d$, so that the magnitude
of $\mu$ may be related to the SUSY breaking scale,  we impose the condition
\beq \label{mu} h+\bar{h}\neq 2\alpha ~. \eeq
A $\mu$--parameter of the right order is induced through the
Guidice--Masiero mechanism  \cite{gm}
via the Lagrangian term, ${\cal L} \supset \int d^{4}\theta
H_{u}H_d\frac{Z^{*}}{M_{pl}}$.  Invariance of this term under the $U(1)$
symmetry requires
\beq \label{gmu} h+\bar{h}-z=0~. \eeq
The gaugino masses arise through the Lagrangian term
\beq \label{gaugino} \int d^{2}\theta
W_{\alpha}W^{\alpha}\frac{Z}{M_{Pl}}~\eeq
once $\left\langle F_Z\right\rangle \neq 0$ is induced.  Combining the
invariance of this term with that of the gauge kinetic term
$\int d^{2}\theta W_{\alpha}W^{\alpha}$, we see that the spurion field
$Z$ must have zero charge under the $U(1)$.\footnote{SUSY breaking scalar masses
are invariant under the $U(1)$ symmetry and do not provide any constraint.}
It is clear that the
simultaneous presence of the gaugino
mass term and the $\mu$ term reduces the $U(1)$ symmetry to a discrete
subgroup $Z_N$.  Therefore, one
has to start with a discrete symmetry $Z_{N}$ with the spurion
superfield $Z$ having a
charge $0 ~mod~N$. Under $Z_N$, the conditions
Eqs. (\ref{mu})-(\ref{gaugino}) become
\begin{eqnarray}
\label{zn}
z&=&0~~mod~N\nonumber\\
h+\bar{h}&=&0~~mod~N\nonumber \\
h+\bar{h}&\neq&2\alpha~~mod~N
\end{eqnarray}
which also implies that $2\alpha\neq 0~mod~N$.

Based on the invariance of the Yukawa couplings of Eq. (\ref{superpotential})
under $Z_n$ and the conditions listed in Eq. (\ref{zn}), we obtain the following
set of constraint equations:
\begin{eqnarray} \label{c}
z&=&m_{1}N\nonumber\\
h+\bar{h}&=&m_{2}N\nonumber\\
q+u+h&=&2\alpha+m_{3}N\nonumber\\
q+d+\bar{h}&=&2\alpha+m_{4}N\nonumber\\
l+e+\bar{h}&=&2\alpha+m_{5}N\nonumber\\
2n&=&2\alpha+m_{6}N\nonumber\\
l+n+h&=&2\alpha+m_{7}N,
\end{eqnarray}
where $m_{i}$ ($i=1-7$) are all integers.

The discrete $Z_N$ anomaly coefficients for the $SU(3)_C$ and the $SU(2)_L$
gauge groups are
\begin{eqnarray}
A_{3} &=&-3\alpha+3q+\frac{3}{2}u+\frac{3}{2}d,\nonumber\\
A_{2}&=&-5\alpha+\frac{9}2q+\frac{3}{2}l+\frac{1}{2}(h+\bar{h})~.
\end{eqnarray}
Here we note that the fermionic charge of the $u^c$ field, for example,
relevant for
the anomaly coefficient, is $(u-\alpha)$ since $\theta$ carries charge
$\alpha$.  $A_3$ and $A_2$ include contributions from the gauginos as
well.  We shall cancel these anomalies by applying the
Green--Schwarz mechanism as given in  Eq. (6).

Non--zero gauginos masses arise through the VEV $\left\langle F_Z \right\rangle
\neq 0$ (see Eq. (11)).  Let us denote the $Z_N$ charge
of $F_Z$ to be $M$.
$\left\langle F_Z \right \rangle \neq 0$ breaks the original $Z_N$ symmetry
down to a subgroup $Z_M$:
\beq Z_{N}\rightarrow Z_{M} ~.
\eeq
(It must be that
$M > 1$ for an unbroken discrete symmetry to survive after SUSY breaking.)
Since $M = z - 2\alpha = m_1N - 2 \alpha$ where $m_1$ is an integer, we
have
\begin{eqnarray}
\alpha = {n_1 \over 2} M
\end{eqnarray}
with $n_1$ being an integer.  Let $N = n_0 M$ where $n_0$ is an integer.
Since invariance of the Yukawa couplings under the $Z_N$ symmetry requires
invariance under the subgroup $Z_M$, we can solve the constraints
of Eqs. (13)-(14) along with Eq. (6)
to determine the various charges by first confining to the
invariance under the smaller group $Z_M$.  Under this $Z_M$, a superpotential
term $\mu H_u H_d$ will be allowed.
Once a solution is found, we can embed the
$Z_M$ symmetry into a higher $Z_N$ symmetry that would forbid the
$\mu$ term.  Making some change of
variables, viz., $n_2 = n_0 m_2,~n_4 = n_0 m_6,~n_5 = -n_0 p_1,~n_6 = -n_0p_2$,
and applying the anomaly cancellation condition of Eq. (6), we obtain the
charges of the various fields from Eqs. (13)-(14) as
\begin{eqnarray} \label{solution}
z&=&Mn_{0}\nonumber\\
h&=&3q+M\left(\frac{n_{2}-n_6}3-\frac{n_{4}}2-\frac{7n_{1}}6\right)+\frac{M}b(n_{2}-n_{3}+
\frac{n_{5}}3-n_{1})\nonumber\\
\bar{h}&=&-3q+M\left(\frac{2n_{2}+n_6}3+\frac{n_{4}}2+\frac{n_{1}}6\right)+\frac{M}b(-n_{2}
+n_{3}-\frac{n_{5}}3+n_{1})\nonumber\\
u&=&-4q+M\left(-\frac{n_{2}-n_6}3+\frac{n_{4}}2+\frac{n_{1}}6\right)+\frac{M}b(-n_{2}+n_{3}
-\frac{n_{5}}3+n_{1})\nonumber\\
d&=&2q+M\left(\frac{n_{2}-n_6}3-\frac{n_{4}}2-\frac{7n_{1}}6\right)+\frac{M}b(n_{2}-n_{3}+\frac{n_{5}}3-n_{1})\nonumber\\
l&=&-3q+M\left(-\frac{n_{2}-n_6}3+\frac{2n_{1}}3\right)+\frac{M}b(-n_{2}+n_{3}-\frac{n_{5}}
3+n_{1})\nonumber\\
e&=&6q+M\left(-\frac{n_{2}-n_6}3-\frac{n_{4}}2-\frac{5n_{1}}6\right)+\frac{2M}b(n_{2}-n_{
3}+\frac{n_{5}}3-n_{1})\nonumber\\
n&=&M\left(\frac{n_{4}+n_{1}}2\right)\nonumber\\
\alpha&=&M\frac{n_{1}}2~.
\end{eqnarray}
Here we have defined $b\equiv k_{3}/k_{2}$.  The $n_i$ in Eq. (17) are
all integers.
A specific choice of the integers $n_i$ will fix the charge assignment
explicitly.  We note that the terms proportional to $q$ in Eq. (17) are
proportional to the SM hypercharge $Y$.  One can remove these terms and
set $q=0$ in Eq. (17) without loss of generality
by making a shift of the $Z_M$ charges
proportional to $Y$. The quark doublet $Q$ will then have
zero charge under the unbroken $Z_M$.  It should
be kept in mind that to each solution we find, one can add $Z_M$ charges
proportional to $Y$ to obtain equivalent solutions.

Based on Eq. (\ref{solution}), one can compute the anomaly coefficients
under $Z_M$.  They are
\begin{eqnarray}
A_3&=&\frac{3}{2}M(n_1-n_2)\nonumber\\
A_2&=&\frac{1}{2b}M(3n_1-3n_2+bn_6)~.
\end{eqnarray}

Note that from the last of Eq. (17), we have $2 \alpha = 0~mod ~M$.
So the superpotential is invariant under $Z_M$.  Also, under $Z_M$, one has
$h+\bar{h} = 0~mod~M$, so a $\mu$ term in the superpotential
is allowed by this symmetry.  (Such a term will be forbidden when the
$Z_M$ symmetry is embedded into a higher $Z_N$ symmetry, which we
shall do in  subsection 3.3.)  In order to avoid
$R$--parity breaking couplings, the total charges of the
corresponding superpotential terms should be non--integer  multiples of $M$, which
puts extra constraints on the $Z_M$ charges.
There are four types of $R$--parity violating terms.  Their
$Z_M$ charges are given by
\begin{eqnarray}
u^{c}d^{c}d^{c}&:&M\left(\frac{5n_{1}}6-\frac{2n_{2}+n_6}3-\frac{n_{4}}2\right)+\frac{M}b(n
_{2}-n_{3}+\frac{n_{5}}3-n_{1})\nonumber\\
LLe^{c}&:&M\left(\frac{n_{1}-n_{4}}2\right)\nonumber\\
LH_u&:&M\left(\frac{n_{1}-n_{4}}2\right)\nonumber\\
QLd^{c}&:&M\left(\frac{n_{1}-n_{4}}2\right).
\end{eqnarray}

It is easy to show  that the largest $Z_M$ symmetry is $Z_{6k_3}$ from
Eq. (17).  We shall now find solutions to these sets of equations for
various values of the parameter $b = k_3/k_2$.

\subsection{Green--Schwarz anomaly cancellation at Kac--Moody level 1}

The simplest possibility for the parameters
$k_2$ and $k_3$ to take is  $k_3 = k_2 = 1$, so that $b=1$ in Eqs. (17)-(18).
This is the case of Kac--Moody algebra realized at level 1.  Since the constraint
equations depend only on the ratio $b=k_3/k_2$, the case of higher levels
will coincide with that of level 1 as long as the levels are the same
for  both $SU(3)_C$ and $SU(2)_L$.  From a theoretical point of view
this case is the most attractive,
since it allows for both $SU(3)_C$ and $SU(2)_L$ to emerge from the same
gauge group as in a GUT.  The charge assignment and possible discrete
symmetries for this case $k_2 = k_3$ are shown explicitly in
Table \ref{b=1} \footnote{Among
the solutions, we remove those which either are conjugate of the
listed solution or can be realized as linear combination of the
known solution and hypercharge. For example, in model III and
IV, we have chosen $q=0~~mod~6$. The charge $q$ need not be actually zero.
Since there exists an unbroken $U(1)_Y$
hypercharge which is anomaly--free, one can always take a linear
combination of model III (or IV) with $U(1)_Y$ to find equivalent
solutions with $q \neq 0~~mod~M$. This comment also applies to the models
listed in all
the other tables.}.

The procedure we have followed to obtain Table 3 is as follows.  First we set $b=1$.
Then we choose a set of integers $n_i$ in the range $n_i \subset (0-5)$.
Any $n_i$ larger than or equal to 6 (or any negative $n_i$) can be absorbed
into the $mod~M$ piece.  The highest $Z_M$ symmetry is then found to be $Z_6$.
For every choice of the integer set $n_i$ we demand that
the $R$--parity breaking couplings
of Eq. (19) be forbidden.  (This requies $n_1-n_4$ to be an odd integer
and that the charge of $d^c$ should be non--zero under $Z_M$.)  Then we
solve for the charges of the various fields, setting $q=0$ as mentioned
above.  If the Green--Schwarz anomaly cancellation condition is satisfied
we accept the solution.  Upto overall conjugation and shifts proportional
to hypercharge, the complete set of solutions is as ginven in Table 3.
We have also listed the anomaly coefficients $(A_2,A_3)$ in Table 3.

\begin{table}[h]
 \begin{center}
  {\renewcommand{\arraystretch}{1.1}
 \begin{tabular}{|c| c |c c c c c c c c c |c| }
   \hline
  \rule[5mm]{0mm}{0pt}Model & $Z_M$ & $q$ & $u$ & $d$ & $l$ & $e$ & $n$ & $h$ & $\bar{h}$ & $\alpha$ &
  $(A_2,~A_3)$ \\
  \hline
  \rule[5mm]{0mm}{0pt}
  I & $Z_2$& 2 & 1 & 1 & 2 & 1 & 1 & 1 & 1 & 1 &  (2, 2) \\
   \rule[5mm]{0mm}{0pt}
  II & $Z_2$&2 & 1 & 1 & 2 & 1 & 1 & 1 & 1 & 0 & (1, 1) \\
   \rule[5mm]{0mm}{0pt}
  III & $Z_6$&6 & 5 & 1 & 2 & 5 & 3 & 1 & 5 & 3 &  (6, 6)\\
   \rule[5mm]{0mm}{0pt}
  IV & $Z_6$&6 & 5 & 1 & 2 & 5 & 3 & 1 & 5 & 0 &  (3, 3) \\
    \hline
\end{tabular}
 }
  \caption{\footnotesize MSSM charge assignment when $k_2=k_3$. $\alpha$
  denotes the charge of the gaugino and the Grassmanian variable $\theta$.
  When $\alpha \neq 0$, the $Z_M$ acts as an $R$--symmetry.
   Also shown are the anomaly coefficients $(A_2,~A_3)$.}
  \label{b=1}
 \end{center}
\end{table}

Several remarks are in order about the results shown in Table 3.

(i) Models I and II differ only in the value of $\alpha$.  In the effective
low energy
Lagrangian, what matters is $2\alpha$, which is the same for both models.
Although the two models look identical from low energy point of view,
their embedding into a high scale theory will not be the same.
This is the reason for listing them separately.  We shall
see that when $Z_M$ is embedded into a higher symmetry $Z_N$ so as to
forbid a large $\mu$ term, Models I and II will look different. Similar
remarks apply to Models III and IV.

(ii) The $Z_2$ symmetries in Table 3 (I and II)
are actually subgroups of the $Z_6$
symmetry (III and IV).
Their embedding into $Z_N$ will however lead to different solutions.
Note also that the $Z_3$ subgroup of $Z_6$ does not show up as a solution,
since that would allow for lepton number violation.

(iii) The $Z_6$ symmetric solutions of Table 3 are actually identical
to the $Z_6$ subgroup of $B-L$ shown in Table 1 which can prevent
$R$--parity violation in MSSM \cite{martin}.  This can be recognized
by taking linear combinations of Models III-IV and hypercharge.  Suppose
we normalize hypercharge so that all MSSM fields have integer values
denoted by $\hat{Y}$ (with $Q$ field having $\hat{Y} = 1$).  Take now the combination
$3(IV) + \hat{Y}$ ($mod~6)$.  This redefined charge is identical
to the $Z_6$ subgroup of $B-L$ of Table 1.  The $Z_2$ models are identified
as the $Z_2$ subgroups of $B-L$. We conclude from our systematic
analysis that even with GS anomaly cancellation, the only allowed
discrete symmetries at the $Z_M$ level (which admits a superpotential
$\mu$ term) are the subgroups of $B-L$.  This
will however be not the case when $Z_M$ is embedded into $Z_N$.  Note also
that the anomaly coefficients $A_2$ and $A_3$ in Table 3 are all equal to
$M/2$, so GS mechanism is not playing a role in anomaly cancellations.
This remark will also be different in the $Z_N$ embedding.

\subsection{Embedding $Z_M$ into $Z_N$ and solving the $\mu$--problem}

We recall that the original $Z_N$ symmetry broke down to a subgroup
$Z_M$ once the spurion field $Z$ acquired a VEV along its $F$--component.
At the level of $Z_M$, a superpotential $\mu$--term is allowed.  Now
we turn to the task of identifying the original $Z_N$ symmetry needed
for explaining the $\mu$ term.  We look for the simplest higher symmetry
into which the $Z_M$ solutions of Table 3 can be embedded.  Each of the
model in Table 3 has a different embedding into $Z_N$.

Consider the embedding of Model I in Table 3 into $Z_N$.  The smallest
$Z_N$ group that contains a $Z_2$ subgroup is $Z_4$.  This embedding is
shown in Table 4.  There are two possible charge assignments indicated
as Models Ia and Ib.  These models are obtained as follows.  First we
choose the value of $\alpha$ to be either 1 or 3 under $Z_4$ (since
it must reduce to 1 under $Z_2$).  These two values correspond to the
two models in Table 4.  Then we demand that a bare $\mu$ term in the
superpotential is prevented by the $Z_4$ symmetry.  That determines
the charges $(h,~\bar{h})$ to be either (1, 3) or (3, 1).  It turns
out that the charges in the latter case are the conjugates of the
former, and so we discard it.  Then we set the charge $n$ to be either
1 or 3, consistent with it being 1 under $Z_2$ subgroup.
This fixes the charges of all fields. For each case
the anomaly coefficients $A_2$ and $A_3$ are computed.  If the anomalies
are cancelled by the GS mechanism, we accept the solution.  Only two
solutions are found to survive, as displayed in Table 4.

\begin{table}[h]
 \begin{center}
  {\renewcommand{\arraystretch}{1.1}
 \begin{tabular}{|c| c c c c c c c c c |c|  }
   \hline
  \rule[5mm]{0mm}{0pt} $Z_4$ & $q$ & $u$ & $d$ & $l$ & $e$ & $n$ & $h$ & $\bar{h}$ & $\alpha$ &
  $(A_2,~A_3)$ \\
  \hline
  \rule[5mm]{0mm}{0pt}
  Ia & 4 & 1 & 3 & 4 & 3 & 1 & 1 & 3 & 1 &  (1, 3)\\
   \rule[5mm]{0mm}{0pt}
  Ib & 4 & 1 & 3 & 4 & 3 & 1 & 1 & 3 & 3 &  (3, 1)\\
        \hline
\end{tabular}
 }
  \caption{\footnotesize
  Embedding of the $Z_2$ symmetry of  Model I of Table 3 into $Z_4$ symmetry. }\label{z4}
 \end{center}
\end{table}

Note that the discrete $Z_4$ anomalies are cancelled by the GS mechanism.
Individually $A_2$ and $A_3$ are not multiples of $N/2 = 2$, but the
two coefficients differ only by $N/2 = 2$.  We conclude that this simple
solution would not have been possible without GS anomaly cancellation.

The models of Table 4 can be recast in a very simple form by forming the
linear combination \{Ia $+ \hat{Y}$ ($mod~4$)\}, or \{Ib $+ \hat{Y}$ ($mod~4$)\}
with $\hat{Y}$ being the integer values of SM hypercharge.  We display
in Table 5 Model Ia recast in this form.  The charge assignment is very
simple, all matter fields of MSSM carry charge 1 under $Z_4$, while the
Higgs superfields carry charge 0.  The gauginos also have charge 1.
The contribution to the $Z_4$ anomaly from the matter fields are the
same for $A_2$ and $A_3$ (the number of color triplets is the same as
the number of $SU(2)_L$ doublets in MSSM).  While the gluino contributes
an amount equal to 3 to $A_3$, the sum of the $SU(2)_L$ gaugino  (= 2)
and the Higgsinos ($= -1$) add to $A_2 = +1$.  We see that $A_2$ and
$A_3$ differ by $N/2 = 2$, signaling anomaly cancellation via GS mechanism.

\begin{table}[h]
 \begin{center}
  {\renewcommand{\arraystretch}{1.1}
 \begin{tabular}{|c| c c c c c c c c c |c|  }
   \hline
  \rule[5mm]{0mm}{0pt} $Z_4$ & $q$ & $u$ & $d$ & $l$ & $e$ & $n$ & $h$ & $\bar{h}$ & $\alpha$ &
  $(A_2,~A_3)$ \\
  \hline
  \rule[5mm]{0mm}{0pt}
  Ia & 1 & 1 & 1 & 1 & 1 & 1 & 0 & 0 & 1 & (3, 1)\\
        \hline
\end{tabular}
 }
  \caption{\footnotesize Model 1a recast with a shift proportional to
  hypercharge.}\label{z4p}
 \end{center}
\end{table}

The charge assignment shown in Table 5 is clearly compatible with grand
unification.  The Kac--Moody level associated with hypercharge will be
$k_1 = 5/3$ with a GUT embedding.  Gauge coupling unification is then
predicted, since $\sin^2\theta_W = 3/8$ near the string scale. This
is true even if there were no covering GUT symmetry.

Now we turn to Model II of Table 3.  Embedding this $Z_2$ into a $Z_4$
is not viable, since a large $\mu$ term cannot be prevented in that
case. The next simplest possibility is $Z_6$, which also does not work
as the $Z_6$ anomalies do not cancel.  The simplest embedding is found to be
into a $Z_{10}$ with the charge assignment as shown in Table 6.

\begin{table}[h]
 \begin{center}
  {\renewcommand{\arraystretch}{1.1}
 \begin{tabular}{|c|  c c c c c c c c c |c | }
   \hline
  \rule[5mm]{0mm}{0pt} $Z_{10}$ & $q$ & $u$ & $d$ & $l$ & $e$ & $n$ & $h$ & $\bar{h}$ & $\alpha$ &
 $(A_2,~A_3)$ \\
  \hline
  \rule[5mm]{0mm}{0pt}
  IIa&10&1&7&4&3&7&3&7&2& (6, 6)\\
   \rule[5mm]{0mm}{0pt}
   IIb&10&7&9&8&1&9&1&9&4& (2, 2)\\
        \hline
\end{tabular}
 }
  \caption{\footnotesize $Z_{10}$ embedding of Model II.}
 \end{center}
\end{table}

In the models of Table 6, one
might consider a $Z_5$ subgroup of $Z_{10}$. This subgroup
is sufficient to forbid the $\mu$--term in the superpotential $W$,
as well as to prevent dangerous $R$--parity violating couplings
in $W$.  With invariance only under $Z_5$, the term
$u^c d^c d^c$ will have zero charge.  A  Lagrangian term arising from
the Kahler potential
${\cal L} \supset \int d^4\theta
(u^c d^c d^c Z^*/M_{\rm Pl}^2)$ will then be allowed.  Once $F_Z$ acquires
a non--zero VEV, this term will lead to a superpotential term
${\cal L} \supset \int d^2\theta (M_{\rm SUSY}/M_{\rm Pl}) u^c d^c d^c$.  Such a term
violates $R$--parity, although very weakly.  Signals of such a weak
violation will be unobservable in collider experiments.  However,
this scenario will not fit well with cosmological constraints.
The LSP will decay through this induced $R$--parity violating Yukawa
coupling $\lambda^{''}$ , which has a strength of order $M_{\rm SUSY}/M_{\rm Pl} \sim
10^{-15}$.  We can estimate the lifetime of the LSP to be
$\tau \sim [(\lambda^{''})^2 M_{LSP}/(8 \pi)]^{-1} \sim 10^4$ sec.
Such a lifetime falls into the cosmologically disfavored range
and would be in violation of nucleosynthesis constraints.  (This
situation is analogous to the gravitino problem of supergravity,
but is slightly worse, since the LSP mass is expected to be order
100 GeV, rather than a TeV for the gravitino,
making the LSP lifetime somewhat longer than that of the gravitino.)
We consider the $Z_5$ solution to be unacceptable for this reason.
Since $Z_5$ symmetry does not contain a $Z_2$ subgroup, exact $R$--parity
could not be defined after SUSY breaking, unlike in the case of $Z_{10}$
model.

In Tables 7 and 8 we show  the simplest embedding of
Models III and IV into $Z_{12}$ and $Z_{18}$ respectively.
The procedure we have adopted is identical to that for Models I and II.

\begin{table}[h]
 \begin{center}
  {\renewcommand{\arraystretch}{1.1}
 \begin{tabular}{|c | c c c c c c c c c| c | }
   \hline
  \rule[5mm]{0mm}{0pt} $Z_{12}$ & $q$ & $u$ & $d$ & $l$ & $e$ & $n$ & $h$ & $\bar{h}$ & $\alpha$ &
  $(A_2,~A_3)$ \\
  \hline
    \rule[5mm]{0mm}{0pt}
  IIIa & 12 & 5 & 7 & 8 & 11 & 9 & 1 & 11 & 3 & (9, 9) \\
   \rule[5mm]{0mm}{0pt}
  IIIb & 12 & 11 & 1 & 8 & 5 & 3 & 7 & 5 & 3 & (9, 9) \\
     \rule[5mm]{0mm}{0pt}
  IIIc & 12 & 5 & 7 & 8 & 11 & 9 & 1 & 11 & 9 & (3, 3) \\
   \rule[5mm]{0mm}{0pt}
  IIId & 12 & 11 & 1 & 8 & 5 & 3 & 7 & 5 & 9 & (3, 3) \\
    \hline
\end{tabular}
 }
  \caption{\footnotesize $Z_{12}$ embedding of Model III.}
 \end{center}
\end{table}

\begin{table}[h]
 \begin{center}
  {\renewcommand{\arraystretch}{1.1}
 \begin{tabular}{|c | c c c c c c c c c |c | }
   \hline
  \rule[5mm]{0mm}{0pt} $Z_{18}$ & $q$ & $u$ & $d$ & $l$ & $e$ & $n$ & $h$ & $\bar{h}$ & $\alpha$ &
  $(A_2,~A_3)$\\
  \hline
  \rule[5mm]{0mm}{0pt}
  IVa & 18 & 11 &  13&14  &17 & 15 & 1 &17  & 6 & (9, 9) \\
  \rule[5mm]{0mm}{0pt}
  IVb  & 18 & 5 &  1& 8 & 11 & 15 &  7& 11 & 6 & (18, 9) \\
       \hline
\end{tabular}
 }
  \caption{\footnotesize $Z_{18}$ embedding of model IV.}
 \end{center}
\end{table}

As in the case of the $Z_{10}$ model of Table 6, we may consider taking
a $Z_9$ subgroup of $Z_{18}$ in Table 8.  However, since $Z_9$ does not
contain $Z_2$ or $Z_6$ as subgroups, after SUSY breaking, small $R$--parity
violating Yukawa couplings of the type $W \supset L L e^c$ will be
generated from the Kahler potential with  coupling constants of
order $10^{-15}$.  Such  couplings would violate constraints from
big bang nucleosynthesis since the lifetime of the LSP will be of order
$10^4$ sec.  We shall not consider  the $Z_9$ subgroup any further.

\subsection{Discrete anomaly cancellation at higher Kac--Moody level}

Thus far we have assumed the parameter $b \equiv k_3/k_2 =1$.  This is the
case when the  Kac--Moody levels for $SU(3)_C$ and $SU(2)_L$
are the same, the simplest possibility being $k_3 = k_2 = 1$.
It is also possible that $k_2$ and $k_3$ are not the same.  It is
not clear to us how easy it is to construct string models
with different values for $k_3$ and $k_2$.  Although it might appear less
attractive theoretically, it is nevertheless a logical possibility.
In this section we analyze discrete anomaly cancellation for values
of $k_3/k_2 \neq 1$.

From a technical point of view it appears to be difficult to construct models
with levels higher than 3 in string theory \cite{dienes}.  Motivated
by this observation, we shall confine our
discussions to $k_2$ and $k_3$ being less than or equal to 3.  This allows for the
cases when $b\equiv k_3/k_2 = 1, 2, 1/2, 1/3, 2/3$ and $3/2$.  The case of $b=1$ has
already been analyzed in the previous section, so we turn to the other cases.

From the solution Eq. (\ref{solution}) which applies to $Z_M$ invariance
(that allows a bare $\mu$ term, but forbids all $R$--parity violations),
a few simplifications can be found.
The case where $b = 1/2$ and $b= 1/3$ are identical to the case of $b=1$.
This is because the $b$--dependent terms only contribute to the various
charges proportional to $n_5$ in Eq. (17).  But this $n_5$ contribution
can be absorbed into the $n_2$ term in all equations.  No new solutions
will then be generated under $Z_M$.
Similarly, it is easy to see that the cases $b=2/1$
and $b=2/3$ are equivalent under $Z_M$.  And the case where $b=3/1$ becomes
identical to the case of $b=3/2$.  Among these equivalent cases under $Z_M$,
we shall only consider one possibility.  Although it is possible that when
the resulting models are embedded into a higher symmetry $Z_N$, new models
at higher levels may emerge, we shall not pursue it here.

We shall then focus on
the case where $b \equiv k_3/k_2=2$, and $b=3$ for anomaly cancellation
at higher Kac--Moody level.
Following the same procedure as in the previous section, we obtain the
corresponding discrete symmetry and charge assignment.  The solutions
for the case of $k_3/k_2 = 2$, which is the same for $k_3/k_2 =2/3$,
are shown in Table 9.  The
discrete $Z_M$ symmetry is $Z_6$ in this case.  Note that the discrete
GS mechanism cancels the gauge anomalies of $Z_6$.  For example,
$A_2 = 9/2,~A_3=6$ is anomaly free since with $k_2=1,~k_3=2$, the
cancellation condition is that $2 A_2$ and $A_3$ differ by
an integer multiple of $N/2 = 3$ (see Eq. (6)).

\begin{table}[hp]
 \begin{center}
  {\renewcommand{\arraystretch}{1.1}
 \begin{tabular}{|c| c |c c c c c c c c c |c| }
   \hline
  \rule[5mm]{0mm}{0pt}
  Model &$Z_M$& $q$ & $u$ & $d$ & $l$ & $e$ & $n$ & $h$ & $\bar{h}$ & $\alpha$
  &$(A_2,~A_3)$ \\
  \hline
 V &$Z_{6}$ & 6 & 2 & 4 & 5 & 5 & 3 & 4 & 2 & 3 & (9/2, 6) \\
 \rule[5mm]{0mm}{0pt}
 VI & $Z_{6}$&6 & 2 & 4 & 5 & 5 & 3 & 4 & 2 & 0  & (3/2, 3)\\
 \hline
\end{tabular}
 }
  \caption{\footnotesize Discrete symmetries and the corresponding charge
  assignment when $k_3/k_2=2$ or $2/3$.}
  \label{b=2}
 \end{center}
\end{table}

\begin{table}[hp]
 \begin{center}
  {\renewcommand{\arraystretch}{1.1}
 \begin{tabular}{|c|  c c c c c c c c c |c | }
   \hline
  \rule[5mm]{0mm}{0pt} $Z_{12}$ & $q$ & $u$ & $d$ & $l$ & $e$ & $n$ & $h$ & $\bar{h}$ & $\alpha$ &
  $(A_2,~A_3)$ \\
  \hline
  \rule[5mm]{0mm}{0pt}
   Va& 12 & 8 & 4 & 5 & 11 & 3 & 10 & 2 & 3 & (9/2, 9)\\
   \rule[5mm]{0mm}{0pt}
   Vb& 12 & 2 & 10 & 5 & 5 & 9 & 4 & 8 & 3 & (9/2, 9) \\
   \rule[5mm]{0mm}{0pt}
   Vc & 12 & 2 & 10 & 11 & 11 & 3 & 4 & 8 & 3 & (3/2, 9) \\
   \rule[5mm]{0mm}{0pt}
   Vd & 12 & 8 & 4 & 11 & 5 & 9 & 10 & 2 & 3 & (3/2, 9) \\
  \hline
\end{tabular}
 }
  \caption{\footnotesize Embedding of $Z_6$ of Model V into $Z_{12}$.}
 \end{center}
\end{table}

\begin{table}[hp]
 \begin{center}
  {\renewcommand{\arraystretch}{1.1}
 \begin{tabular}{|c | c c c c c c c c c |c | }
   \hline
  \rule[5mm]{0mm}{0pt} $Z_{18}$ & $q$ & $u$ & $d$ & $l$ & $e$ & $n$ & $h$ & $\bar{h}$ & $\alpha$ &
  $(A_2,~A_3)$\\
  \hline
  \rule[5mm]{0mm}{0pt}
  VIa& 18 & 2 &  10&11  &5 & 15 & 4 &14  & 6 &  (9/2, 18)\\

  \rule[5mm]{0mm}{0pt}
   VIb & 18 & 14 &  16& 5 & 17 & 15 &  10& 8 & 6 & (27/2, 9) \\
     \hline
\end{tabular}
 }
  \caption{\footnotesize $Z_{18}$--embedding of the $Z_{6}$ model VI.}
 \end{center}
\end{table}

The two $Z_6$ models of Table 9 have been embedded into the simplest possible
$Z_N$ model in Tables 10 and 11.  The $Z_N$ symmetries are found to
be $Z_{12}$ and $Z_{18}$. The discrete gauge anomalies are cancelled
by GS mechanism, as before.  Take Model Vd for example, which has
$A_2=3/2,~A_3=9$ under $Z_{12}$.  $2A_2$ and $A_3$ differ by 6,
which is an integer multiple of $N/2 = 6$.

The next case is when $b = k_3/k_2 = 3$, which gives at the $Z_M$ level the
same models as $b=3/2$.  In Table 12 we list the allowed $Z_M$ models, with
$M= 18$.
Table 13 has the embedding of Model VII into $Z_{18}$ that prevents
a large $\mu$ term, Table 14 has the embedding of Model VIII, the simplest possibility
for which being $Z_{90}$.  In all cases the discrete anomalies are cancelled
by the GS mechanism.

\begin{table}[hp]
 \begin{center}
  {\renewcommand{\arraystretch}{1.1}
\begin{tabular}{| c |c| c c c c c c c c c  |c | }
   \hline
  \rule[5mm]{0mm}{0pt} Model & $Z_M$ &$q$ & $u$ & $d$ & $l$ & $e$ & $n$ & $h$ & $\bar{h}$ & $\alpha$ &
  $(A_2,~A_3)$ \\
  \hline
   \rule[5mm]{0mm}{0pt}
   VII& $Z_{18}$&18 & 1 & 17 & 10 & 7 & 9 & 17 & 1 & 9 & (6,18)\\
    \rule[5mm]{0mm}{0pt}
   VIII&$Z_{18}$& 18 & 1 & 17 & 10 & 7 & 9 & 17 & 1 & 18 & (15, 9)\\
     \hline
\end{tabular}
}
  \caption{\footnotesize Discrete symmetries and the charge assignments
  for $k_3/k_2 = 3$ or $3/2$.}
  \label{b=3}
 \end{center}
\end{table}

\begin{table}[hp]
 \begin{center}
  {\renewcommand{\arraystretch}{1.1}
 \begin{tabular}{|c |c c c c c c c c c |c | }
   \hline
  \rule[5mm]{0mm}{0pt} $Z_{36}$ & $q$ & $u$ & $d$ & $l$ & $e$ & $n$ & $h$ & $\bar{h}$ & $\alpha$ &
  $(A_2,~A_3)$ \\
  \hline
  \rule[5mm]{0mm}{0pt}
  VIIa  & 36 & 1 &  35& 28 & 25 & 9  &  17& 19 & 9 & (33, 27)\\
  VIIb  & 36 & 19 &  17& 10 & 7 & 9  &  35& 1 & 9 & (6, 27)\\
   \hline
\end{tabular}
 }
  \caption{\footnotesize $Z_{36}$--embedding of the $Z_{18}$ model VII.}
 \end{center}
\end{table}

\begin{table}[hp]
 \begin{center}
  {\renewcommand{\arraystretch}{1.1}
 \begin{tabular}{|c |c c c c c c c c  c |c | }
   \hline
  \rule[5mm]{0mm}{0pt} $Z_{90}$ & $q$ & $u$ & $d$ & $l$ & $e$ & $n$ & $h$ & $\bar{h}$ & $\alpha$ &
  $(A_2,~A_3)$ \\
  \hline
  \rule[5mm]{0mm}{0pt}
  VIIIa  & 90 & 55 &  71& 46 & 25 & 9 &  53& 37 & 54 & (89, 27) \\
   \rule[5mm]{0mm}{0pt}
  VIIIb  &  90& 55 &  53& 28 & 25 & 27 & 89 &1  &  72& (42, 36)  \\
    \hline
\end{tabular}
 }
  \caption{\footnotesize Embedding of $Z_{18}$ of Model VIII into $Z_{90}$.}
 \end{center}
\end{table}

It should be mentioned that at the level of $Z_M$, it is easy to realize
an $R$--parity that allows for lepton number violation, but conserves
baryon number.  Rapid proton decay will be prevented in this case.
Lepton number violating processes and neutrino masses do provide some constraints,
but these are much less stringent.

Consider the $Z_6$ models of Table 3 (Models III and IV).  Suppose we
impose invariance only under the $Z_3$ subgroup of $Z_6$ with the same
charge assignment as in Table 3.  The lepton number violating couplings
$LH_u,~LLe^c,$ and $QLd^c$ all have charge $3$ under the $Z_6$
symmetry, so with only $Z_3$ invariance imposed, these couplings will
be allowed in the superpotential.  Since the original $Z_6$ symmetry
is free from discrete gauge anomalies, the subgroup $Z_3$ is also free
from such anomalies.  One cannot however embed this $Z_3$ symmetry to any
higher $Z_N$ in order to explain the $\mu$--parameter.  Consider the
$LH_u$ term in the superpotential.  $Z_N$ invariance of this term would
imply $l+h = 2\alpha~mod~N$.  The last two relations of Eq. (13)
would imply $2\alpha = 0~mod~N$, implying that a bare $\mu$ term in
the superpotential will be allowed.  An alternative explanation for
the $\mu$--term will have to be found in the case of lepton number
violating $R$--parity.

It is also possible, although somewhat non--trivial, to have baryon
number violating $R$--parity without dangerous lepton number violation.
(Neutrino masses violate lepton number by two units, but that
does not result in rapid proton decay.)
At the level of $Z_N$ we can show that anomalies associated with
such an $R$--parity will have to be cancelled at higher Kac--Moody levels.
If the coupling $u^c d^c d^c$ is allowed in the superpotential, we find
that the $Z_N$ discrete symmetry has anomalies given by $A_3 = 3 \alpha$,
$A_2 = 5/2 \alpha - (3/4) pN$ where $p$ is an integer. Imposing the anomaly
cancellation condition, Eq. (6), we find $\alpha = (m_1k_2 +
m_2-(k_3/2)p)N/(6k_3-5k_3)$ with $m_1,m_2,p$ being integers.  When $k_2=k_3=1$,
this relation shows that $2\alpha = 0 ~mod ~N$, meaning that a bare
$\mu$--term will be allowed in the superpotential.  If we choose $k_2$
and $k_3$ differently, this problem will not arise. Consider for example,
$k_3=2, k_2=1$.  A consistent $Z_N$ charge assignment corresponding to
a $Z_4$ symmetry is shown in Table 15 for this case.
This model allows for the coupling
$u^cd^cd^c$, while preventing other $R$--parity violating couplings.
The $Z_4$ anomalies are cancelled by GS mechanism, which in this case
reads as $2A_2-A_3 = m-2n$, with $m,n$ being integers.

\begin{table}[hp]
 \begin{center}
  {\renewcommand{\arraystretch}{1.1}
 \begin{tabular}{|c |c c c c c c c c  c |c | }
   \hline
  \rule[5mm]{0mm}{0pt} $Z_{4}$ & $q$ & $u$ & $d$ & $l$ & $e$ & $n$ & $h$ & $\bar{h}$ & $\alpha$ &
  $(A_2,~A_3)$ \\
  \hline
  \rule[5mm]{0mm}{0pt}
  A  & 4 & 2 &  2 & 1 & 1 & 1 &  2 & 2 & 1 & (1/2, 3) \\
   \rule[5mm]{0mm}{0pt}
  B  &  0& 2 &  2& 3 & 3 & 3 & 0 & 0 &  3& (5/2, 5 )  \\
    \hline
\end{tabular}
 }
  \caption{\footnotesize Examples of a $Z_4$ symmetry that allows for
  baryon number violation without dangerous lepton number violations.
  The $Z_4$ anomalies are cancelled by GS mechanism at levels $k_2=1, k_3=2$.}
 \end{center}
\end{table}

\section{Discrete flavor symmetries and the the fermion mass hierarchy}

As indicated earlier, there must be an unbroken $Z_M$ symmetry
which is flavor--independent that survives to low energy scales,
to be identified as an $R$--parity.  We can however introduce
flavor dependence in the original symmetry, provided that a
subgroup of the flavor group remains unbroken and can be
identified as one of the $Z_M$ symmetries of Table 3.  In
this section we embark on this question.  Our aim will be
to seek an understanding of the observed hierarchy in the
fermion masses and mixings without introducing such hierarchy
by hand.

Anomalous $U(1)_A$ symmetry is widely used  for the explanation of
fermion mass and mixing hierarchy \cite{fn}. The general
superpotential in this has the following expression:
\begin{eqnarray}
 \label{mssm1}
 W&=&
Q_i u^{c}_j H_u \left(\frac{S}{M_{\rm Pl}}\right)^{{(h_1)}_{ij}} +
Q_id^{c}_jH_d\left(\frac{S}{M_{\rm Pl}}\right)^{{(h_2)}_{ij}} + L_i
e^{c}_jH_d
\left(\frac{S}{M_{\rm Pl}}\right)^{{(h_3)}_{ij}} \nonumber \\
&+& L_i\nu^{c}_jH_u\left(\frac{S}{M_{\rm Pl}}\right)^{{(h_4)}_{ij}}+
\nu^{c}\nu^{c}S\left(\frac{S}{M_{\rm Pl}}\right)^{{(h_5)}_{ij}}
\end{eqnarray}
where $S$ is an MSSM singlet field with a non--trivial anomalous
$U(1)_A$ charge. $S$ acquires a VEV near the string scale and
disappears from the low energy spectrum. Here
$(h_{\alpha})_{ij}$ is a set of integers for
$\alpha=1,2,3,4,5$ and $i,j=1,2,3$ are the generation indices.
We assume that all the Yukawa couplings
are of order one. After $S$ field develops a VEV,  near but somewhat
below the string scale, a small parameter $\epsilon =
\left\langle S \right\rangle/M_{\rm Pl} \sim 1/5$ is generated.  This
factor appears in various powers with the Yukawa couplings, explaining
the observed mass and mixing hierarchy \cite{fn}.
It is possible to suppress all the MSSM Yukawa couplings to the desired
level by choosing appropriate set of $U(1)$ charges \cite{mass}.

An acceptable flavor texture which gives the correct pattern of
fermion masses and mixings is:
\begin{eqnarray}
\label{matr1}
 U_{ij}&=&\left(
\begin{array}{ccc}
  \epsilon^{6} & \epsilon^{5} & \epsilon^{3} \\
  \epsilon^{5} & \epsilon^{4} & \epsilon^{2} \\
  \epsilon^{3} & \epsilon^{2} & 1 \\
\end{array}
\right)H_{u}, ~~~~ D_{ij}=\left(
\begin{array}{ccc}
  \epsilon^{4} & \epsilon^{3} & \epsilon^{3} \\
  \epsilon^{3} & \epsilon^{2} & \epsilon^{2} \\
  \epsilon & 1 & 1 \\
\end{array}
\right)\epsilon^{p}H_{d},\nonumber\\
L_{ij}&=&\left(
\begin{array}{ccc}
  \epsilon^{4} & \epsilon^{3} & \epsilon \\
  \epsilon^{3} & \epsilon^{2} & 1 \\
  \epsilon^{3} & \epsilon^{2} & 1 \\
\end{array}
\right)\epsilon^{p}H_{d}, ~~~~~ \nu^D_{ij}=\left(
\begin{array}{ccc}
  \epsilon^{2} & \epsilon & \epsilon \\
  \epsilon & 1 & 1 \\
  \epsilon & 1 & 1 \\
\end{array}
\right)\epsilon^{a_1}H_{d},
\end{eqnarray}
where $U_{ij}$, $D_{ij}$, $ L_{ij}$ and $\nu^D_{ij}$ correspond to
up--quark, down quark, charged lepton and Dirac neutrino Yukawa matrices
resulting from the appropriate powers of the  $S$ field in Eq. (\ref{mssm1}).
The integer $p$ can be either  0, 1 or 2, corresponding to large, medium and small
$\tan \beta$ respectively.

Once the charged lepton sector and Dirac neutrino sector are
constructed, we can uniquely define the form of
the heavy Majorana neutrino mass matrix.  In the
present example it is
 \beq
 \label{neu4}
\nu_{ij}^M=\left(
\begin{array}{ccc}
  \epsilon^{2} & \epsilon & \epsilon \\
  \epsilon & 1 & 1 \\
  \epsilon & 1 & 1 \\
\end{array}
\right)\epsilon^{a_2}~. \eeq

As mentioned before, the MSSM superpotential does not possess any
unbroken $U(1)$ symmetry, apart from the gauge symmetry.
Therefore we seek solutions of a $Z_N$ discrete symmetry that
would generate the Yukawa matrices of Eq. (21).

$Z_N$ invariance of the Yukawa
couplings in Eq. (\ref{mssm1}) imposes  constaints
in the up--quark sector given by
\beq \label{qq1}
 q_i + u_j + h+ ps = 2\alpha~mod~N,
 \eeq
where $p$ is the power of $\epsilon$ appearing in the appropriate
element of the $U_{ij}$ matrix,  which is equal to the power in the
field $S$.  $s$ denotes the $U(1)$ charge of $S$ field.
Similar conditions apply for the charges of the other SM fermions as
well.  By construction, the flavor structure of the matrices obey
the ``determinant rule", viz., that in any $2 \times 2$ sub--block,
the determinant is a  homogeneous function of $\epsilon$.  This means
that out of the 18 conditions for the up--quark and the down--quark
sector, only 8 will be independent.

We wish to have an unbroken $Z_M$ symmetry that is a subgroup of
$Z_N$ which is flavor--independent.  Since the flavon field $S$
has a $Z_N$ charge of $s$, once it acquires a VEV of order the
string scale, the $Z_N$ will be broken down to $Z_s$.  We shall
attempt to embed the $Z_M$ of Table 3 into $Z_s$.

To be specific, let us work out an example with the $Z_2$ model of
Table 3 and embed this $Z_2$ into a higher $Z_N$ symmetry that
allows for the desired flavor structure.  The $Z_N$ symmetry
must be $Z_{14+2n}$ for this embedding to be consistent, with
$n$ being any integer.  The smallest such symmetry is then
$Z_{14}$.  The flavon field $S$ must carry zero charge under
$Z_2$ and should  transform non--trivially under the $Z_N$.
The simplest possibility is $s=2$.  Now, the Yukawa textures of
Eq. (21) makes use of $S^6$ terms in the superpotential, which should
be different from $S^0$.  This requirement makes the smallest
$Z_N$ symmetry to be $Z_{14}$.  If this symmetry were $Z_{12}$,
for example, $S^6$ will be neutral under $Z_{12}$, making the
(11) and the (33) entries of the up--quark Yukawa matrix of the
same order.  We can
generalize this statement to any low scale discrete symmetry.  The
corresponding flavor--dependent symmetry must be $Z_{M(k+1)+n}$, where
$Z_{M}$ is the low scale surviving discrete symmetry, $k$
corresponds the highest power of the $S$ field in the general superpotential,
Eq. (\ref{mssm1}), and $n$ is a positive integer. This
choice will guarantee the existence of $Z_M$ discrete $R$--parity
at low energy scales.

Three examples of $Z_{14}$ symmetric models are presented in
Table 16.  We have chosen the charge of $S$ to be 2 and fixed the
charge of $\theta$ to be 7 in these examples.  Discrete anomaly
cancellation is enforced via GS mechanism at Kac--Moody level 1.
We have also imposed the conditions that the $Z_{14}$ symmetry
forbid all $R$--parity violating couplings.

\begin{table}
{\small
 \begin{center}
  {\renewcommand{\arraystretch}{1.1}
  \begin{tabular}{ |c| c| c| c|  c|  c|  c|  c|  c|  c|  c|c  c| }
     \hline
   \rule[5mm]{0mm}{0pt}
  &$Q_i$&$u^{c}_i$&
   $d^{c}_i$&$L_i$&
   $e^{c}_i$&$\nu^{c}_i$ &
   $H_{u}$&$H_{d}$&
   $\theta$&$S$&$A_2$&$A_3$\\
   \hline
   \rule[5mm]{0mm}{0pt}
    A&0,2,6&1,3,7&3,5,5&4,6,6&13,1,5&5,7,7&1&13&7&2 &6& 13\\
    \rule[5mm]{0mm}{0pt}
    B&4,6,10&13,1,5&11,13,13& 6,8,8&9,11,1&5,7,7&13&1&7&2&13&13\\
    \rule[5mm]{0mm}{0pt}
    C&6,8,12&5,7,11&1,3,3&0,2,2&7,9,13&5,7,7&9&5&7&2&13&6\\
   \hline
  \end{tabular}
  }
  \caption{\footnotesize Examples of flavor--dependent $Z_{14}$ symmetry which
  forbids all $R-$parity breaking terms.
   $i=1,2,3$ is the flavor index and charges in the brackets are in order of 1-3.
   We are considering $p=2$
   and $q=0$ in Eq. (\ref{matr1}) which  corresponds to medium
   values of $\tan\beta\sim 10$. We have taken $a_2=0$ in Eq.
   (\ref{neu4})
   for simplicity.}

 \end{center}}
\end{table}

The $Z_N$ symmetry group would depend on the highest power of $\epsilon$
appearing in the fermion Yukawa matrices. If we want to have symmetry smaller
than $Z_{14}$,  we should reduce the power of $S$ field in
Eq. (\ref{mssm1}).  One way is to re-parameterize the
value of $\epsilon$. For example, if $\epsilon$ is taken to be of
order 1/10, rather than 1/5 as was assumed in Eq. (21), it might  suffice
to use cubic powers of $S$ at most.  A $Z_8$ discrete symmetry would then
suffice to forbid the $R$--parity breaking terms.  We consider the
expansion given in Eq. (21) to be more realistic.

In Table 17 we present three models based on $Z_{28}$ symmetry
that forbid all $R$--parity violating couplings,  explain the
fermion mass and mixing hierarchy via the texture of Eq. (21)
and also solve the $\mu$--problem via the Guidice--Masiero
mechanism.  As before, the discrete gauge anomalies are cancelled
by the GS mechanism.  We find it remarkable that a single
discrete symmetry can do all these jobs.  It may be mentioned that
$Z_{28}$ is not a large symmetry unlikely to be realized in string
theory.  For example, if the particle spectrum contains fields carrying
charges of $(1,1/4,1/7)$ under the anomalous $U(1)$, and if a scalar field
with charge 1 acquires a VEV, the unbroken
$Z_N$ symmetry will be $Z_{28}$.

 \begin{table}{\small
 \begin{center}
  {\renewcommand{\arraystretch}{1.1}
  \begin{tabular}{ |c| c| c| c|  c | c|  c|  c|  c|  c|  c| c  c|}
     \hline
   \rule[5mm]{0mm}{0pt}
  &$Q_i$&$u^{c}_i$&$d^{c}_i$&$L_i$&$e^{c}_i$&$\nu^{c}_i$&$H_{u}$&$H_{d}$&$\theta$&$S$&$A_2$&$A_3$\\
   \hline
   \rule[5mm]{0mm}{0pt}
   A&12,16,24&7,11,19&9,13,13&4,8,8&17,21,1&3,7,7&27&1&7&4&11&11\\
    \rule[5mm]{0mm}{0pt}
   B&22,26,6&23,27,7&1,5,5&2,6,6&21,25,5&3,7,7&1&27&7&4 &11&11\\
    \rule[5mm]{0mm}{0pt}
   C&26,2,10&7,11,19&9,13,13&18,22,22&17,21,1&3,7,7&13&15&7&4&11&25\\
   \hline
  \end{tabular}
  }
  \caption{\footnotesize
  Examples of flavor--depended discrete  $Z_{28}$ symmetry
   which  prevent $R$--parity breaking couplings, explain the origin of
   the $\mu$--term via the Guidice--Masiero mechanism, and explain the
   hierarchy in quark and lepton masses and mixings via the Yukawa texture
   shown in Eq. (21).
   $i=1,2,3$ is the flavor index and charges in the brackets are in order of 1-3.
   We   are considering $p=2$
   and $q=0$ in Eq. (\ref{matr1}), which corresponds to medium values
   of  $\tan\beta\sim 10$. We  have taken $a_2=0$ in Eq.
   (\ref{neu4}) for simplicity. }
  \label{MSSM-33}
 \end{center}}
\end{table}

\section{Conclusion}

In this paper we have investigated the possibility of realizing $R$--parity
of MSSM as a discrete gauge symmetry.  Simultaneously we have demanded that
this discrete symmetry should provide a natural explanation for the
$\mu$--term, the Higgsino mass parameter in the MSSM superpotential, via
the Guidice--Masiero mechanism.
We have adopted a discrete version of the Green--Schwarz anomaly cancellation
mechanism in our search for discrete gauge symmetries, which is
is less constraining than the conventional methods.

We have found simple examples of $Z_N$ symmetries that act as $R$--parity
and simultaneously solve the $\mu$--problem, without extending the
particle content of the MSSM.  The simplest example
is a $Z_4$ symmetry with a simple charge assignment that is compatible
with grand unification.  The Green--Schwarz mechanism plays a crucial
role in cancelling the $Z_4$ anomalies.  Other examples of $Z_N$
symmetries are provided with $N=10,12,18,36$ etc.  In some cases
the discrete anomalies are cancelled by the GS mechanism at higher
Kac--Moody levels.  We have found that it is easy to realize lepton number
violating $R$--parity as a discrete symmetry, but implementing the
Guidice--Masiero mechanism for the $\mu$ term is difficult in this case.
Baryon number violating $R$--parity can be realized, along with a natural
explanation of the $\mu$ term, but the discrete gauge anomalies are cancelled
in this case at higher Kac--Moody level.

It has been shown that a simple $Z_N$ symmetry can also explain the observed
hierarchical structure of quark and lepton masses and mixings, while
preserving the desired $R$--parity and the solution to the $\mu$--problem.
Examples of a $Z_{28}$ symmetry doing all these have been presented.

\section{Acknowledgement}

We thank J. Lykken and Ts. Enkhbat for useful discussion. This
work is supported in part by DOE Grant \# DE-FG03-98ER-41076, a
grant from the Research Corporation and by DOE Grant \#
DE-FG02-01ER-45684.

\end{document}